\documentclass[12pt,halfline,a4paper,]{ouparticle}

\usepackage{hyperref}
\usepackage{graphicx}
\usepackage{listings}
\usepackage{xcolor}
\usepackage{fancyvrb}
\usepackage{framed}

\hypersetup{breaklinks=true,
            bookmarks=true,
            pdfauthor={},
            pdftitle={Tracking in apps' privacy policies}
            }

\usepackage{booktabs}
\usepackage{longtable}

\usepackage{amssymb, amsmath, geometry, amsfonts, verbatim, endnotes, setspace}

\DefineVerbatimEnvironment{Highlighting}{Verbatim}{commandchars=\\\{\}}
\definecolor{shadecolor}{RGB}{248,248,248}

\IfFileExists{upquote.sty}{\usepackage{upquote}}{}

\providecommand{\tightlist}{%
  \setlength{\itemsep}{0pt}\setlength{\parskip}{0pt}}

\makeatletter
\def\Newlabel#1#2#3{\expandafter\gdef\csname #1@#2\endcsname{#3}}

\def\Ref#1#2{\@ifundefined{#1@#2}{???}{\csname #1@#2\endcsname}}

\newcommand*\ifcounter[1]{%
  \ifcsname c@#1\endcsname
    \expandafter\@firstoftwo
  \else
    \expandafter\@secondoftwo
  \fi
}

\usepackage[citestyle=ieee,style=numeric]{biblatex}
\begin{document}

\title{Tracking in apps' privacy policies}

\author{%
%
\name{Konrad Kollnig}\address{University of Oxford}\email{\href{mailto:firstname.lastname@cs.ox.ac.uk}{firstname.lastname@cs.ox.ac.uk}}
%
%
%
%
}

\abstract{Data protection law, including the General Data Protection Regulation (GDPR), usually requires a
privacy policy before data can be collected from individuals. We analysed 15,145 privacy policies from
26,910 mobile apps in May 2019 (about one year after the GDPR came into force), finding that only opening
the policy webpages shares data with third-parties for 48.5\% of policies, potentially violating the GDPR.
We compare this data sharing across countries, payment models (free, in-app-purchases, paid) and platforms (Google Play Store, Apple App Store).
We further contacted 52 developers of apps, which did not provide a privacy policy, and asked them about their data practices.
Despite being legally required to answer such queries, 12 developers (23\%) failed to respond.}

\date{August 2019 (Revised November 2021)}


\maketitle

\hypertarget{introduction}{%
\section{Introduction}\label{introduction}}

App developers often share data with numerous third-parties to serve ads and analyse usage behaviour \autocite{maps_2019,china_2018,reuben_trackers_2018}. This \emph{tracking} produces incredibly rich and valuable data, which can pose risks if in the wrongs hands. This problem with data collection is exacerbated by a centralisation of data by a few tech companies, essentially creating a single point-of-attack.
App users are often not aware that they are being tracked. However, so far, few studies exist that analyse data collection in apps, beyond free Android apps.

At the same time, most data protection and privacy laws, including the GDPR in the EU and UK, implement some form of \emph{notice and consent}, that is, informing the data subject about data practices and asking them for consent.
In this scheme of notice and choice, ideally, a data subject should be able to familiarise themselves with the data protection terms of an app by first reading the privacy policy -- but without data collection taking place -- and only make a decision on whether to give consent, after having read this policy.

In this paper, we analyse whether users are actually informed \emph{before} data collection happens, by focusing on \emph{tracking in apps' privacy policy webpages}. While this does not give insights into the actual data collection within apps, this allows for comparison across devices (Android, iOS), payment models (free, paid), and countries (EU, non-EU).
For a subset of apps without a privacy policy, we additionally contacted the app developers for reasons.

Based on the above observations, this study aims to investigate two, related research questions:

\begin{enumerate}
\def\labelenumi{\arabic{enumi}.}
\tightlist
\item
  To what extent do app developers allow their users to read the relevant privacy policies before collecting personal data about them?
\item
  How does third-party tracking compared across devices (Android, iOS), payment models (free, paid), and countries (EU, non-EU)?
\end{enumerate}

\hypertarget{the-gdpr-and-the-need-for-a-privacy-policy}{%
\section{The GDPR and the need for a privacy policy}\label{the-gdpr-and-the-need-for-a-privacy-policy}}

The GDPR came into effect on 25 May 2018, and imposes high standards for the protection of data that relates to individuals (`personal data').
It applies to all companies that do business in the EU or UK, or process data relating to individuals from these countries. Non-compliance faces very high fines.

Personal data, as protected under the GDPR, reaches beyond obvious identifiers such as name, address, or telephone number. Other means of identification, such as online identifiers, may already suffice to relate data uniquely to an individual, rendering such data personal (Recital 30 GDPR). The same applies to device or advertising identifiers, commonly used for user identification in mobile apps \autocite{dpdb_sar_2018}. Even IP addresses can qualify as personal data, especially if combined with further data \autocite{ecj_ip}.

The GDPR differentiates three main actors: the \emph{data subject}, the \emph{data controller} and the \emph{data processor}. The data controller decides how the personal data of the data subject is processed by the data processor. In the case of mobile apps, typically, the user is the subject, and the app developer the controller -- sometimes jointly with other third-parties that facilitate tracking if such is used.

The GDPR grants data subjects various rights, including the \emph{right to be informed} (Article 12--14). This right obliges data controllers to communicate the terms of personal data processing, at the time such is processed.
For mobile apps, the data controller must typically provide a \emph{privacy policy} before app installation, which should additionally be accessible from within the app \autocite{a29_202}.
Among other aspects, the privacy policy must clarify the purposes of data processing, the legal bases upon which data processing is based, and the recipients of personal data. For a full account of what must be disclosed, see Articles 12--14 GDPR.

An app user has a \emph{right to be informed} about the terms of personal data collection, ``at the time'' of collection (Article 13 GDPR). Such is necessary for the data subject to make an informed decision about the usage of an app or any other service that processes personal data. This view aligns with the judgement of the EU Article 29 Working Party, stating that

\begin{quote}
the data subject must have the necessary information at his disposal in order to form an accurate judgement. In order to avoid any ambiguity, such information must be made available \emph{before} any personal data is processed \autocite[ Section 3.4.1: Consent prior to installation and processing of personal data, emphasis added]{a29_202}
\end{quote}

In the following, we will now analyse whether app users can actually read apps' privacy policies, before being exposed to potentially unwanted data collection.

\hypertarget{methodology}{%
\section{Methodology}\label{methodology}}

In this study, we focus on the \emph{number of tracking parties contacted, when opening the privacy policy URL of an app}. Every app developer has the option to provide such a privacy policy on the app store presence. On iOS, such provision has been compulsory as of 3 October 2018, for every new app or update submitted to the App Store; on the Google Play Store, this obligation has now also been introduced, but after our study.

Focusing on the number of tracking parties in privacy policy pages provides a consistent metric across countries, platforms (iOS and Android) and payment models (free, paid, and in-app purchases). Further, there exist ample tools to detect such web tracking, allowing for scale and representative insights.

We first compiled a representative list of apps on iOS and Android, from different countries (UK, US, Germany, France, Poland) and across different categories (top free, top paid, top grossing, top new free, top new paid) from the online service \texttt{42matters}.
Top grossing included those top free and paid apps, generating most revenue (from direct sales or in-app purchases).
The \emph{top new} and the \emph{top new paid} categories were only available for Android apps. The iOS platform was further divided into app charts for iPad and iPhone apps. Each pair of country and category comprises \(540\) apps on Android and \(1,500\) apps on iPhone and iPad each, leading to a total of \(26,771\) unique apps. The data collection was conducted on 11 May 2019, about one year after the GDPR came into force.

For every one of these apps, we browsed its privacy policy URL, if provided on the app store presence, and counted the number of embedded trackers. The privacy policies were opened with Google Chrome, and the trackers detected with \texttt{Ghostery}, a prominent anti-tracking tool. \texttt{Ghostery} enables Chrome users to detect and block trackers on websites, by detecting network communications to known tracker domains.

We automated this process with the website testing tool \texttt{puppeteer}, that allows to control a Chrome instance from the command line. This allowed to iterate over all policy URLs on our list of apps without supervision. For every policy, we waited waited until all HTML elements of the page were loaded, plus a further 3 seconds to allow the website scripts to load further content from the web.
A modified \texttt{Ghostery} would then store information on all detected trackers to a database.

Other researchers have used dedicated website analysis tools, such as \href{https://phantomjs.org/}{PhantomJS} or \href{https://github.com/mozilla/OpenWPM}{OpenWPM}. Whilst these tools may allow for higher detection rates and better scalability, they often do not reflect the real webrowsing experience of a user and are sometimes sporadically updated, being research software. In fact, PhantomJS is currently not further developed, and is known to render websites differently than end-user browsers. So, we decided to use Google Chrome and \texttt{Ghostery}, to reflect a realistic end-user browsing experience and reliable tracking detection.

Some apps were found not to provide a privacy policy. So, we asked app developers what personal data they process and why they do not provide a privacy notice, under the GDPR's \emph{right to be informed}.
This right was invoked by sending emails to the developers. Developers have one month to answer such requests (Article 12 GDPR).
If we had had not received an answer one week before the deadline, we sent a reminder to the app developer, similar to previous such studies. We selected app developers of the top paid apps, which came more often without privacy policy than free apps. This practice was approved by our departmental ethics board.

\hypertarget{results}{%
\section{Results}\label{results}}

A total number of \(26,771\) unique apps was inspected with regards to tracker in their privacy policy pages.

Of the considered apps, \(3,996\) (15\%) apps provide no privacy policy on the respective app stores, yielding a total of \(22,775\) policy URLs to analyse. \(578\) policy URLs returned error pages, some of which included trackers, and are regarded as valid policy pages, since they may still expose users to unexpected tracking. Two thirds of policies (\(15,552\), 68\%) are transferred using HTTPS encryption. The top three formats to deliver policy are HTML (\(21,741\), 95.5\%), PDF (\(564\), 2.5\%), and plaintext (\(107\), 0.5\%). The top three domains are from Google Sites (\(480\), 2.2\%); Google Docs (\(261\), 1.2\%); and Iubenda, a privacy policy generator (\(241\), 1.1\%). All these website providers may include trackers, such as Google Analytics.

\begin{figure}
\includegraphics[width=1\linewidth]{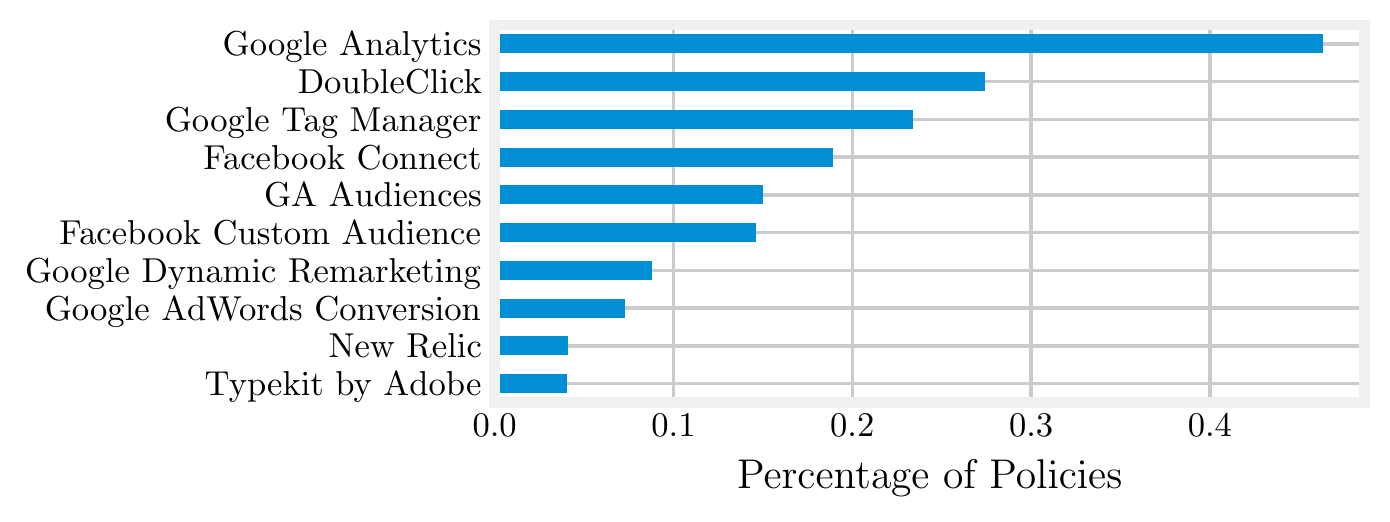} \caption{Top 10 trackers on the $22,775$ policy URLs.}\label{fig:toptrackers}
\end{figure}

\begin{table}
    \centering
    \begin{tabular}{cccccc}
        \toprule
        Count & Std & Mean & Median & 75th & 95th \\ 
        \midrule 
        22\,775 & 5.36 & 2.94 & 1 & 3 & 14  \\
        \bottomrule 
    \end{tabular}
    \caption{Descriptive statistics about trackers in the $22\,775$ policy URLs.}
    \label{tab:trackers_general}
\end{table}

Descriptive statistics about the trackers in the policies are provided in Table \ref{tab:trackers_general}. 56\% of policies contain trackers, 53\% analytics trackers, and 36\% advertising trackers. The top 10 trackers are displayed in Figure \ref{fig:toptrackers}; the top 8 trackers stem from either Google or Facebook.

\begin{figure}
    \begin{subfigure}{0.48\linewidth}
        \centering
        \includegraphics[width=\linewidth]{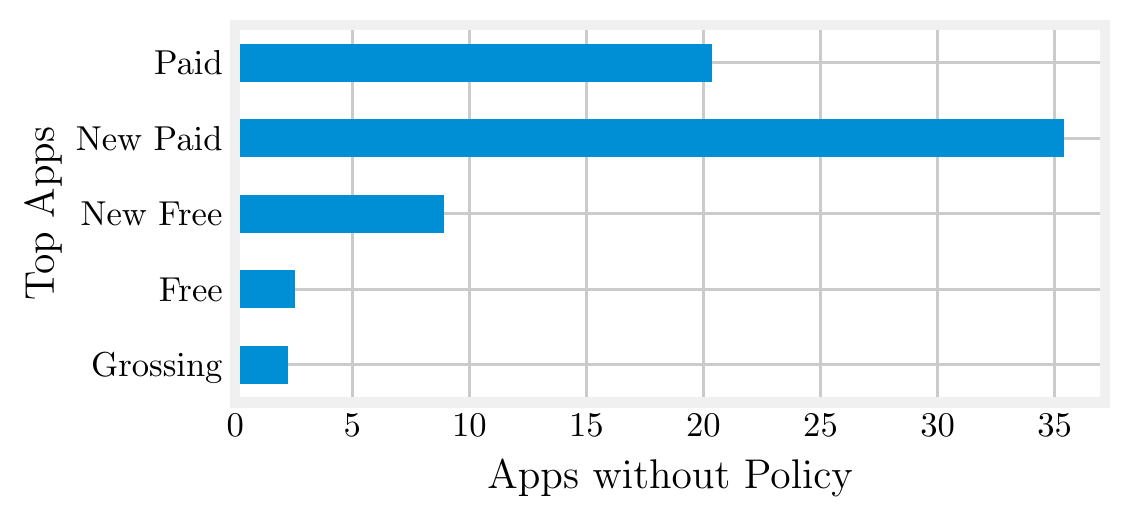}
        \caption{Android Categories.}
        \label{fig:collection_plot_haspolicy}
    \end{subfigure}
    \begin{subfigure}{0.48\linewidth}
        \centering
        \includegraphics[width=\linewidth]{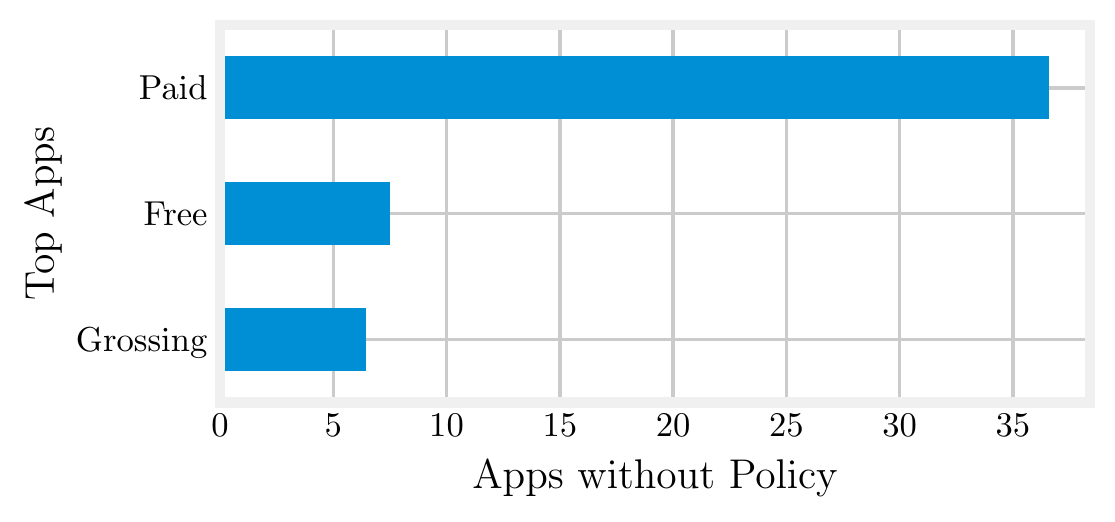}
        \caption{iOS Categories.}
        \label{fig:collection_plot_ios_haspolicy}
    \end{subfigure}
    \begin{subfigure}{0.48\linewidth}
        \centering
        \includegraphics[width=\linewidth]{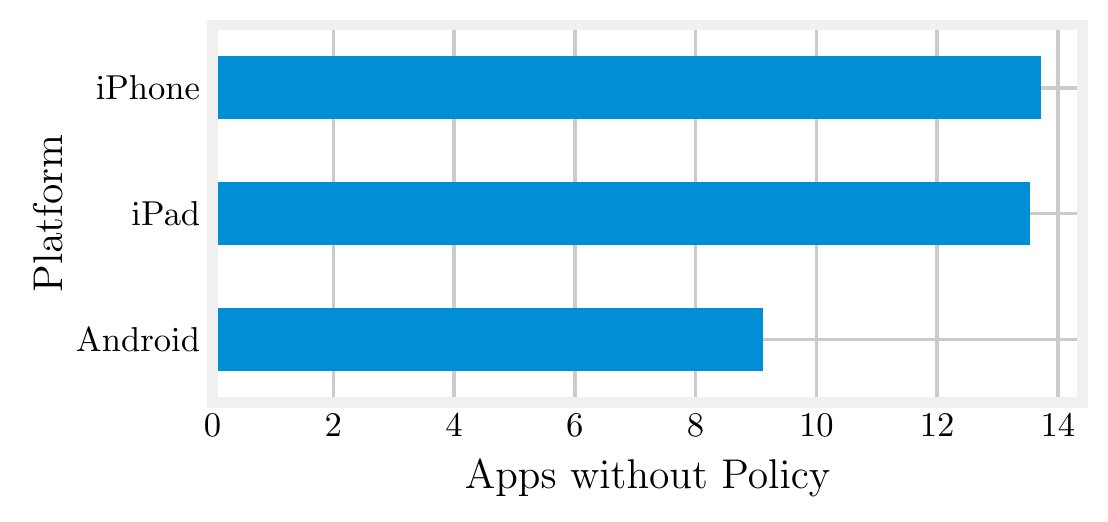}
        \caption{Platform.}
        \label{fig:platform_plot_haspolicy}
    \end{subfigure}
    \begin{subfigure}{0.48\linewidth}
        \centering
        \includegraphics[width=\linewidth]{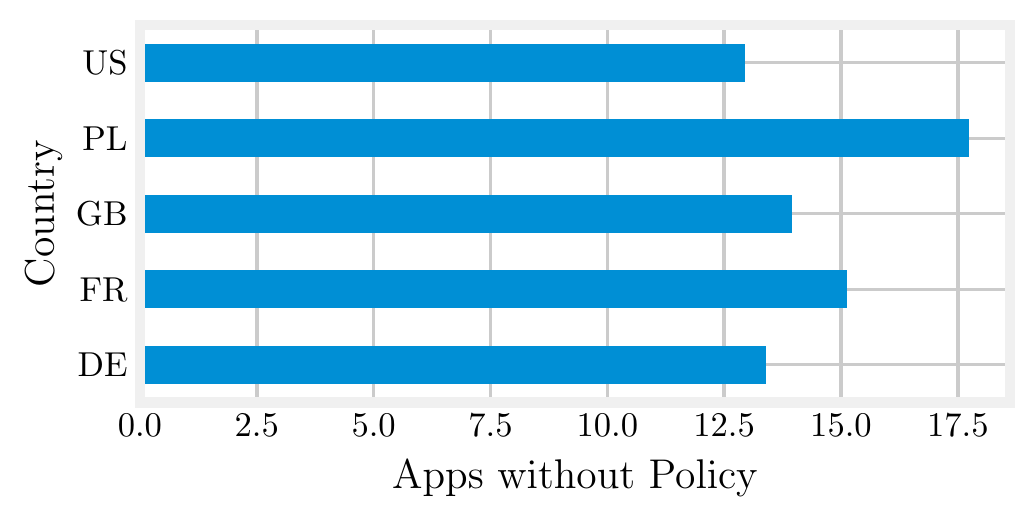}
        \caption{Store countries.}
        \label{fig:country_plot_haspolicy}
    \end{subfigure}
    \caption[Top apps without privacy policy]{Percentage of top apps without privacy policy. Since iOS does not provide statistics about new apps, such are excluded from Figure~\ref{fig:platform_plot_haspolicy}.
    Since there are fewer Android apps than iOS ones in the dataset, only the top 540 apps are considered for each triplet of country, category, and platform, in Figure~\ref{fig:platform_plot_haspolicy}.}
    \label{fig:policies}
\end{figure}

Figure \ref{fig:policies} compares the presence of privacy policies by store category, platform, and store country. Platform-wise, the provision of policies differs, with more Android (9\%) apps providing a policy than iOS (14\%). On both platforms, most \emph{top free} and \emph{top grossing} apps provide a privacy policy, whilst more than a fifth in the two paid categories does not (20\% on Android, 37\% on iOS). The same holds for new apps, which come less often with a privacy policy than \emph{top free} and \emph{top grossing} apps. The US store has least apps without a privacy policy (13\%), whilst Poland has most (18\%).

\begin{figure}
    \begin{subfigure}{\linewidth}
        \centering
        \includegraphics[width=0.9\linewidth]{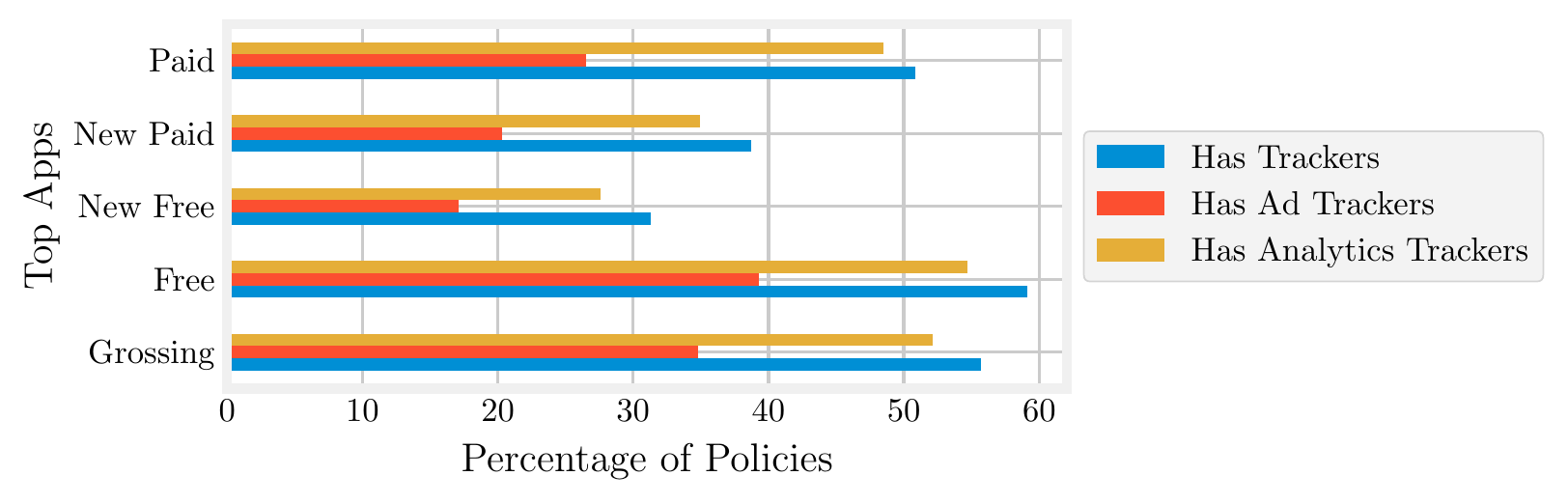}
        \caption{Android Categories.}
        \label{fig:collection_plot}
    \end{subfigure}
    \begin{subfigure}{\linewidth}
        \centering
        \includegraphics[width=0.9\linewidth]{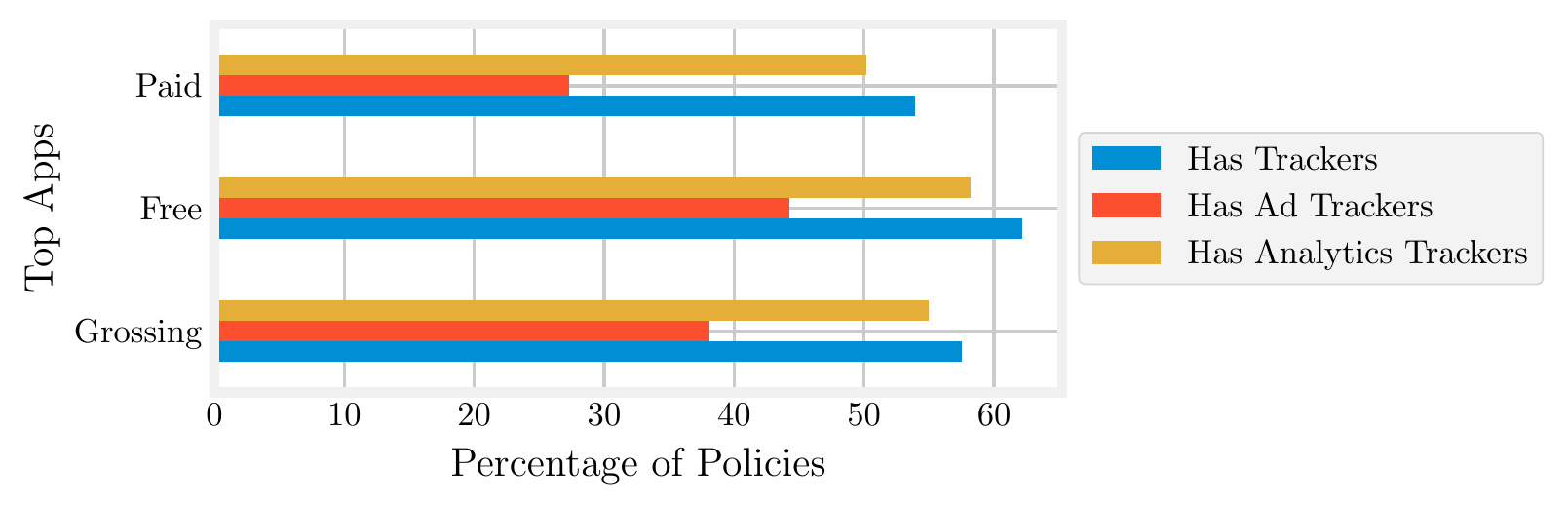}
        \caption{iOS Categories.}
        \label{fig:collection_ios_plot}
    \end{subfigure}
    \begin{subfigure}{\linewidth}
        \centering
        \includegraphics[width=0.9\linewidth]{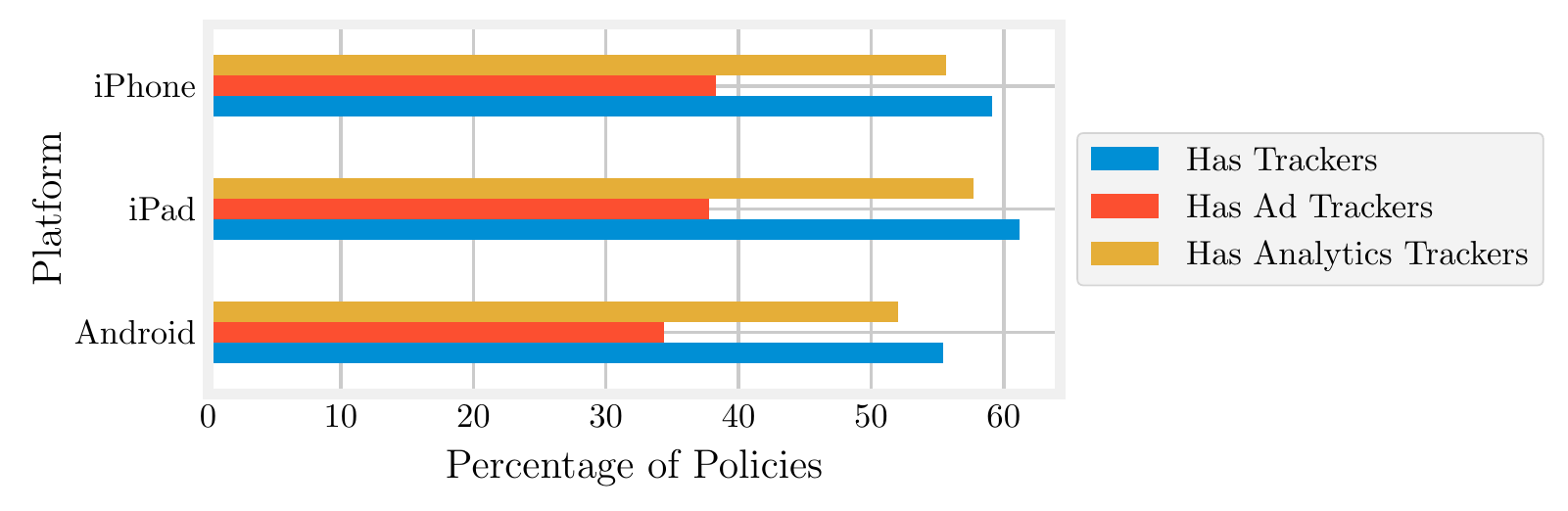}
        \caption{Platforms.}
        \label{fig:platform_plot}
    \end{subfigure}
    \begin{subfigure}{\linewidth}
        \centering
        \includegraphics[width=0.9\linewidth]{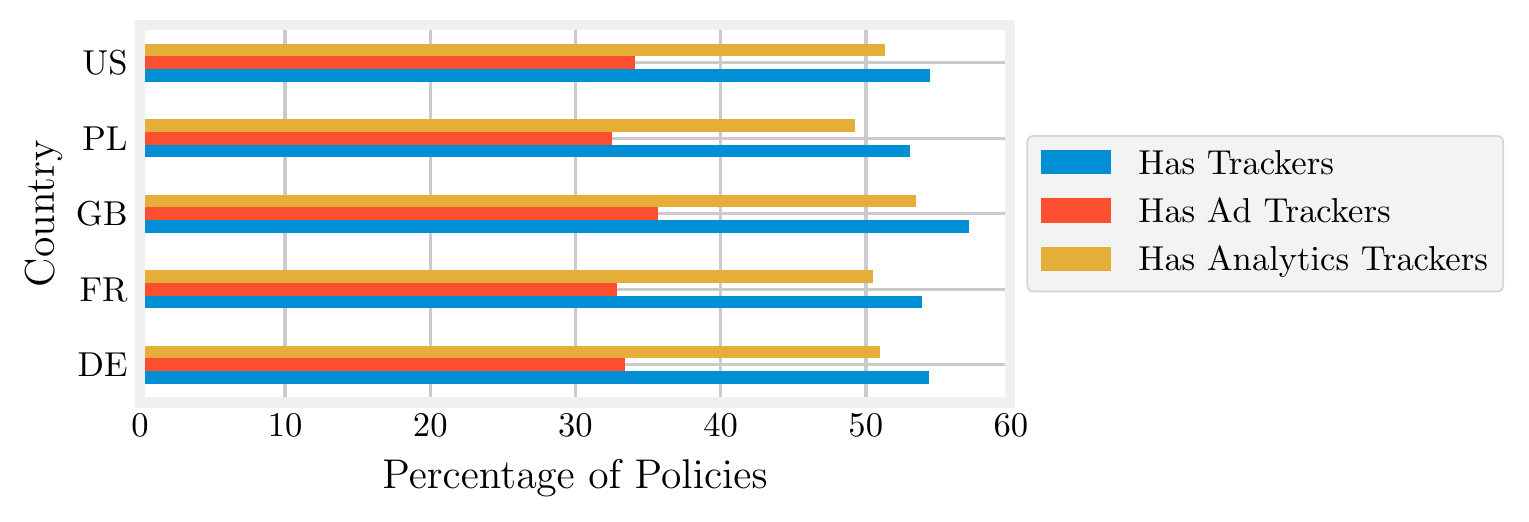}
        \caption{Store Countries.}
        \label{fig:country_plot}
    \end{subfigure}
    \caption[Trackers in apps' privacy policies]{Trackers in the policies 
    of top apps. As in Figure~\ref{fig:policies}, the number of iOS and 
    Android apps are restricted for a fair comparison in Figure~\ref{fig:platform_plot}.}
    \label{fig:trackers}
\end{figure}

Figure \ref{fig:trackers} shows the presence of trackers in the policies, differentiating between analytics and advertising trackers. The presence of policy differs significantly between the categories, see Figure \ref{fig:collection_plot}. \emph{Top free} apps show most trackers, closely followed by \emph{top grossing} apps. The policies of paid apps contain fewer trackers than those of \emph{top free} and \emph{top grossing} ones, but more than of new apps. \emph{New paid} apps come with least trackers, fewer than \emph{new free} apps. Platform-wise, iOS apps show more trackers than Android (56\%), both in analytics and advertising trackers. iPad (61\%) apps show a slightly increased presence of trackers compared to iPhone (59\%) ones. Country-wise, apps on the Polish app store show the smallest presence of trackers (53\%), whereas the British app store has the highest (57\%).

We contacted a total of 52 app developers who did not provide a privacy policy, and asked them about their data practices. Despite being legally required to answer such queries, 12 developers (23\%) failed to respond.

\hypertarget{discussion}{%
\section{Discussion}\label{discussion}}

\textbf{Tracking in privacy policies.} Most app policies contained trackers. A user opening the privacy policy will thus share personal data with a third-party, before having had an opportunity to read and understand the terms of data collection. The top 5\% of policies were even found to contain 14 trackers or more. These observations are in conflict with the right to be informed under GDPR.

\textbf{Use of insecure communication channels.} Only two thirds of privacy policies (68\%) were transferred using HTTPS encryption. By contrast, 87\% of all pages visited through Google Chrome on Android did use HTTPS\autocite{google_https}. Unencrypted communications can easily be intercepted by adversaries. If personal data is transmitted, a lack of encryption may even violate the principle of \emph{integrity and confidentiality} of GDPR (Article 5 GDPR). The upcoming version of Android (API 28) will block unencrypted communication from apps, unless explicitly disabled by the developer.

\textbf{Widespread lack of privacy policy.} There have been identified significant differences in tracking, depending on the store category (e.g.~\emph{top grossing}) of the app. Compared to free apps, paid apps were both more likely to contain fewer trackers in their policy, and to come with no policy at all. This could indicate that paid apps tend to collect less data than free ones, because they do not pursue any data collection to disclose. There remains, however, an obligation for both paid and free apps to provide a privacy policy, if information is accessed or stored on the user's device, under the ePrivacy Directive. This will apply to most apps and highlights another potential violation of the GDPR.

iOS apps were found to provide a privacy policy less often than Android ones. This is despite the previous chapter having outlined that all iOS apps must provide a privacy policy, as opposed to Android ones (although this has been addressed by Google after our data collection). Apple only recently introduced the requirement of privacy policies in October 2018, applying to every new or updated app. iOS apps without a privacy policy have just not been updated on the store since.

\textbf{Limited differences between countries or platforms.} The analysis did not identify major differences between the platforms or countries with regards to tracking in privacy policies. There are two possible reasons for this. First, the policies use the same set of standard trackers, with Google Analytics being the most common one (see Figure \ref{fig:toptrackers}). Second, the set of top apps is usually similar, irrespective of platform or country. Whilst one could analyse the unique apps per platform or country, such an approach would render the resulting set of apps unrepresentative of the day-to-day exposure of users. App users are exposed to a similar level of tracking in privacy policies, independent of country or platform.

\textbf{Paying for apps does not always improve privacy.} Our study also also highlights that paying for apps does not necessarily imply better data protection. \emph{Top free} and \emph{top grossing} apps were both found to contain similar numbers of trackers in their policies. In particular, \emph{top grossing} apps contain a similar number of advertising trackers, despite generating significant revenue from user purchases. This illustrates how even apps, not relying on advertising revenues, may harvest personal data -- especially when they rely on in-app purchases.

\textbf{Many developers did not respond to GDPR requests.} Despite legally binding, we found that many developers did not reply to our GDPR requests, even after having sent them a reminder.
Some of the email addresses that the developers might just not be monitored, so the requests never actually ended up at the developers. However, there will also be emails that the developers received, but did decide not to answer. This points to a lack of understanding and awareness of their legal requirements with developers, as already found in previous studies.

\textbf{Limitations.} This analysis did not regard all apps on the app stores, but only the most popular and recent ones. This is to provide insights into the day-to-day exposure of app users, instead of analysing every app on the app stores. \texttt{Ghostery} relies on a manually curated list of trackers. This means that some trackers will be missed. Moreover, there might be a bias with regards to the physical location of the users of \texttt{Ghostery}, being developed by a German company. For instance, Asian markets are dominated by other tech companies than in the West and the results might differ. The presented analysis does not provide any insight into trackers and GDPR violations within apps. Unfortunately, there currently exist no analysis for a comparison of the program code of free/paid and iOS/Android apps. As a solution, this thesis introduces the first scalable, cross-platform static analysis method in the following section.

\hypertarget{conclusions}{%
\section{Conclusions}\label{conclusions}}

This study analysed tracking of individuals when opening the privacy policy webpages of mobile apps.
We found that most apps share data with third-parties, when a user accesses their privacy policy, which potentially violates the GDPR.
Apps' privacy policies are more often transferred over unencrypted channels than regular websites, another potential GDPR violation. Despite being legally obliged to disclose the storage of data on the user's device, many app developers did not provide a privacy policy, and did not answer to our emails seeking clarification.

As a way forward, we believe that there should be more practical guidance on the technical implications of GDPR. Legal scholars have widely criticised the failure of GDPR to inform the underlying software development process \autocite{dpbd_2017,dpbd_dpbdf_2018}. This is despite the GDPR explicitly prescribing this practice through the principle of Data Protection by Design, under Article 25 of the law.
This paper has provided some technical evidence for this criticism of GDPR.
Good guidance could potentially not only improve compliance with GDPR in practice, but also remove the pressure on the funding of the DPAs.
We have created such a practical guide for app developers at \url{https://gdpr4devs.com}.

\textbf{Future work.} Our study analysed a large subset of popular apps. Follow-up work might consider an even larger dataset, trying to analyse the whole app ecosystem. This is absolutely feasible with our approach due to its immense and easy scalability. There is also a focus on privacy for EU users; follow-up work could analyse the practices in other jurisdictions.
Furthermore, there is a emerging question around what responsibility app platforms themselves should face over apps' adherence to data protection and privacy laws.
We leave this question for other academic studies.

\printbibliography

\end{document}